\documentclass[mathleft
]{an}
\usepackage{graphicx}
\usepackage{rotating}
\usepackage{lscape}
\usepackage{longtable}
\usepackage{times}
\begin{document}

\Pagespan{789}{}
\Yearpublication{2014}%
\Yearsubmission{2014}%
\Month{11}%
\Volume{999}%
\Issue{88}%


\title{The Chinese comet observation in AD 773 January}

\author{J. Chapman\inst{1} \thanks{E-mail: jessechapman@berkeley.edu}, M. Csikszentmihalyi\inst{1}, R. Neuh\"auser\inst{2}}

\titlerunning{The Chinese comet observation in AD 773 January}
\authorrunning{Neuh\"auser \& Neuh\"auser}

\institute{
$^{1}$ Department of East Asian Languages and Cultures, UC Berkeley, Berkeley CA, 94720, United States \\
$^{2}$Astrophysikalisches Institut, Universit\"at Jena, Schillerg\"asschen 2-3, 07745 Jena, Germany}

\received{2014}
\accepted{2014}
\publonline{ }


\abstract
{The strong $^{14}$C increase in the year AD 774/5 detected in one German and two Japanese trees
was recently suggested to have been caused by an impact of a comet onto Earth and
a deposition of large amounts of $^{14}$C into the atmosphere (Liu et al. 2014). 
The authors supported their claim using a report of a historic Chinese observation 
of a comet ostensibly colliding with Earth's atmosphere in AD 773 January.
We show here that the Chinese text presented by those authors is not an original
historic text, but that it is comprised of several different sources.
Moreover, the translation presented in Liu et al. is misleading and inaccurate.
We give the exact Chinese wordings and our English translations.
According to the original sources, the Chinese observed a comet in mid January 773,
but they report neither a collision nor a large coma, just a long tail.
Also, there is no report in any of the source texts about {\em dust rain in the daytime}
as claimed by Liu et al. (2014), but simply a normal dust storm.
Ho (1962) reports sightings of this comet in China on AD 773 Jan 15 and/or 17 and in Japan on AD 773 Jan 20 (Ho 1962).
At the relevant historic time, the Chinese held that comets were produced within the Earth$^{\prime}$s atmosphere,
so that it would have been impossible for them to report a {\em collision} of a comet with Earth$^{\prime}$s atmosphere.
The translation and conclusions made by Liu et al. (2014) are not supported by the historical record.
Therefore, postulating a sudden increase in $^{14}$C in corals off the Chinese coast precisely in mid January 773 (Liu et al. 2014)
is not justified given just the $^{230}$Th dating for AD $783 \pm 14$.}

\maketitle

\section{Introduction: The AD 774/5 event}

Miyake et al. (2012) found a strong increase in the $^{14}$C to $^{12}$C isotope ratio
in two Japanese trees from the year AD 774 to 775.
They excluded supernovae as a possible cause due to the lack
of any historic observations and of any young nearby supernova remnants, and they also excluded
solar super-flares as a cause, because their spectra would not sufficiently explain the
$^{14}$C to $^{10}$Be production ratio observed for that time.
Then, Usoskin \& Kovaltsov (2012), Melott \& Thomas (2012), Thomas et al. (2013), and Usoskin et al. (2013)
suggested that a solar super-flare beamed with only $\ge 24^{\circ}$ degree beam size could have
caused the event (Melott \& Thomas 2012), in particular if four to six times less $^{14}$C was
produced than calculated in Miyake et al. (2012) due to a different carbon circulation model (Usoskin et al. 2013).
Hambaryan \& Neuh\"auser (2013) suggested that a short hard Gamma-Ray-Burst could have caused the event,
because all observables including the $^{14}$C to $^{10}$Be production ratio are consistent with such a burst.
Eichler \& Mordecai (2012) argued that a large solar flare cannot explain the event 
(as also argued in Miyake et al. 2012), but an impact of a massive comet onto the Sun may be able to explain the energetics.

\begin{figure*}
\vspace{-12cm}
\begin{center}
{\includegraphics[angle=0,width=16cm]{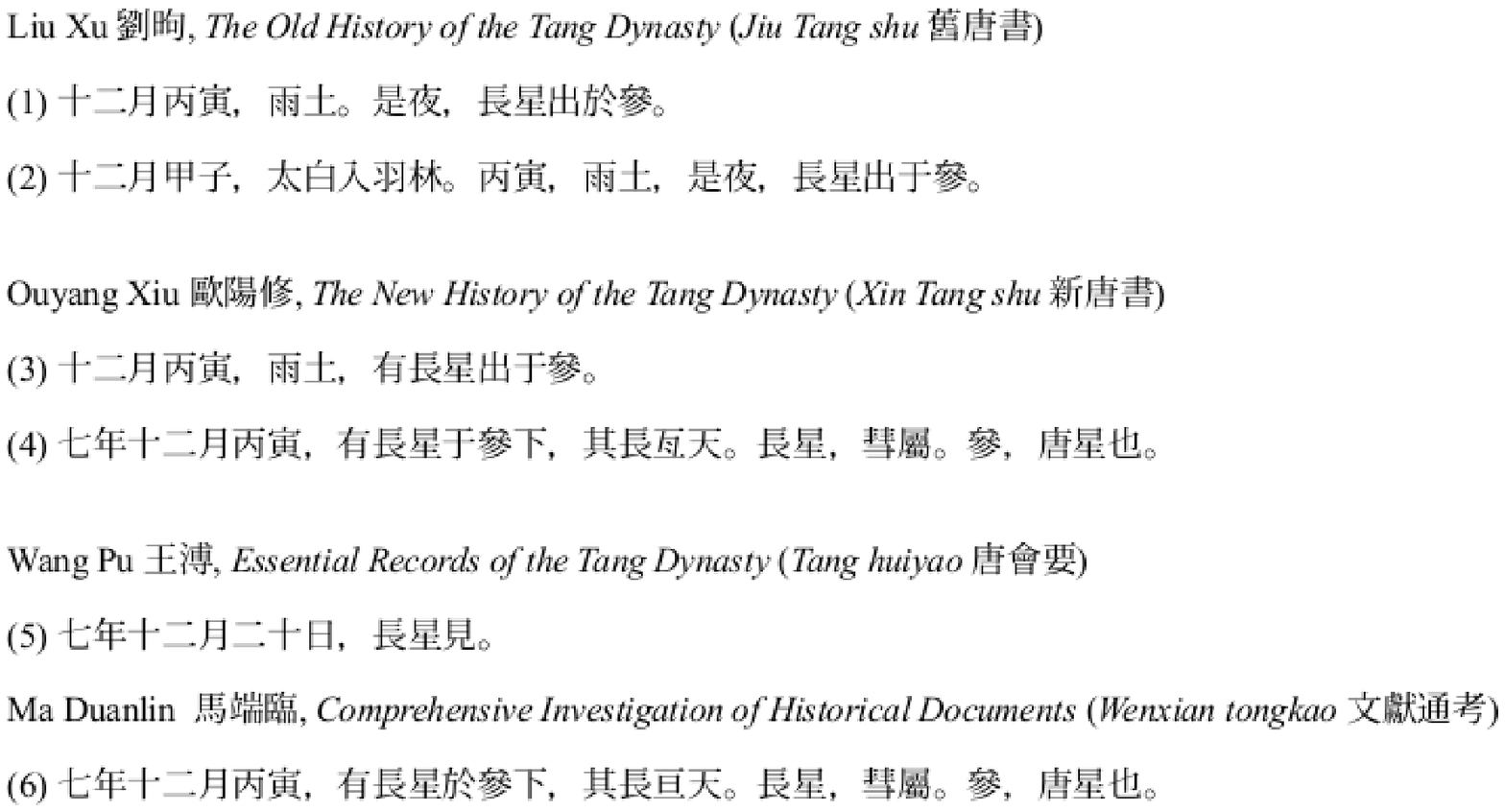}}
\caption{Original Chinese texts about the comet of AD 773 Jan 17,
see Sect. 2 for details and English translation.}
\end{center}
\end{figure*}

Liu et al. (2014) recently obtained additional $^{14}$C measurements of corals off the Chinese coast,
which have a much higher time resolution of, e.g., two weeks, while tree rings have a one-year time resolution.
They found strong variations and a spike in $^{14}$C at around AD $783 \pm 14$ ($^{230}$Th dating),
i.e. possibly near AD 774/5. 
Liu et al. (2014) claim that the first rise in $^{14}$C seen in their data correlates with the
sighting of a comet collision with the Earth$^{\prime}$s atmosphere 
recorded during the Tang dynasty (AD 618-907), on AD 773 Jan 17.
Given this dating, they then conclude that the variations seen in their corals are consistent with the
$^{14}$C variations seen in the Japanese and German trees from AD 774 to 775.

Even more recently, Usoskin \& Kovaltsov (2014) show that such a large amount of $^{14}$C could not be deposited 
in the Earth by a comet nor by an asteroid, unless by a very large body, which would cause severe devastation; 
however, Usoskin \& Kovaltsov (2014) also fall short of questioning the presumed observation presented
by Liu et al. (2014) and whether the Chinese observation really could have been a collision of a comet with Earth.

Therefore, we clarify here the observation 
made by the Chinese in the 12th month of the 7th year of the Dali reign period, AD 773 January.
We present the original Chinese texts and their sources together with our English translations
in Sect. 2 and conclude with our results in Sect. 3.

\section{The original Chinese text}

Liu et al. (2014) present their supposedly historic text in Chinese and English in their figure 2.
Their English translation is as follows: \\
{\em A comet collided with the Earth$^{\prime}$s atmosphere from the constellation of Orion on 17 Jan AD 773 
with coma stretched across the whole sky and disappeared within one day, with $^{\prime}$dust rain$^{\prime}$ in the daytime.} \\
(Liu et al. 2014).

Liu et al. (2014) attribute this quotation to a certain Old Tang Dynasty Book.
This book title is a misleading and inaccurate translation of the title of 
Liu Xu$^{\prime}$s (AD 887-946) {\em The Old History of the Tang Dynasty} ({\em Jiu Tang shu}),
which does not date to the Tang dynasty (AD 618-907), 
but was compiled during the Later Jin (Hou Jin), between AD 940 and 945.

The precise Chinese quotation presented in Liu et al. (2014) is not found in any pre-modern Chinese text.  
Rather, it is an amalgam of several quotations traceable to two sour- ces, namely 
Liu Xu$^{\prime}$s {\em The Old History of the Tang Dynasty} (AD 887-946) and 
Ouyang Xiu$^{\prime}$s (AD 1007-1072) {\em The New History of the Tang Dynasty} (Xin Tang shu; compiled AD 1043-1060).

These two sources relate the events as follows (Chinese texts given in Fig. 1): \\
Those in the earlier history compiled by Liu Xu read: \\
(1) {\em On the bingyin day of the twelfth month (AD 773 Jan 17), a dust storm occurred.
That night, a long star emerged in Shen.} 
\footnote{{\em Jiu Tang shu} (Beijing: Zhonghua, 1975): 11.301.
The asterism Shen corresponds to seven bright stars in Orion: $\alpha$, $\beta$, $\gamma$, $\delta$, $\epsilon$, $\zeta$,
and $\kappa$ Ori.
}  \\
(2) {\em On the jiazi day of the twelfth month (AD 773 Jan 15), Venus entered 
Yulin.\footnote{Yulin is a large asterism containing numerous stars in Austrinus, as well as the northern portion of Piscis Austrinus.}
On the bingyin day (AD 773 Jan 17) a dust storm occurred. That night, a long star emerged in Shen.\footnote{{\em Jiu Tang shu}: 36.1327} 
}

The quotations in the {\em New History} of Ouyang Xiu, compiled AD 1043-1060, read: \\
(3) {\em On the bingyin day of the twelfth month (AD 773 Jan 17), a dust storm occurred, and there was a long star 
that emerged in Shen.\footnote{{\em Xin Tang shu} (Beijing: Zhonghua, 1975): 6.176}} \\ 
(4) {\em On the bingyin day of the twelfth month of the seventh year (AD 773 Jan 17), 
there was a long star beneath Shen.
Its length extended across the sky. 
Long stars belong to the class of comets. Shen is the constellation corresponding 
to the Tang.}\footnote{{\em Xin Tang shu}: 32.838}

Each of the two dynastic histories records the event twice, first in its basic annals, 
which serve primarily as a chronicle of political history, and second in its astronomical treatise.  
The sparser records, quotations number 1 and number 3, belong to the basic annals, 
while the astronomical treatises present the more detailed records, quotations number 2 and number 4.

In addition to the two standard histories of the Tang,
the comet is also mentioned in two medieval Chinese sources
that Liu et al. (2014) do not cite, respectively Wang Pu$^{\prime}$s (AD 922-982) 
{\em Essential Records of the Tang Dynasty} (Tang Huiyao), number 5 below,
and Ma Duanlin$^{\prime}$s (ca. AD 1254 to ca. 1323) 
{\em Comprehensive Investigation of Historical Do- cuments} (Wenxian tongkao), number 6 below. \\
(5) {\em On the twentieth day of the twelfth month of the seventh year 
(AD 773 Jan 17) a long star appeared}.\footnote{{\em Tang hui yao},
3 Vols. (Shanghai: Shangwu, 1935): 2.43.767} \\
(6) See quotation number 4 above; the sole difference between 
these two quotations is a single orthographic variant:
Both texts have characters pronounced {\em gen} and meaning 
{\em extend across}.\footnote{{\em Wenxian tongkao},
2 Vols., (Beijing: Zhonghua, 1986): 1.286.2270b}

This particular comet is also listed in Ho (1962), which is not cited by Liu et al. (2014).    
Ho (1962) gives the date of the comet as AD 773 Jan 15, 
but also remarks that the {\em New History of the Tang} gives the date of AD 773 Jan 17.  
This error is likely due to an eye-skip, 
in which Ho (1962) attributes the {\em jiazi} date of the immediately preceding entry, 
concerning the location of Venus on AD 773 Jan 15, to the appearance of the comet, 
only a few characters later in the text.  
Both the {\em Old} and {\em New History of the Tang Dynasty} give the date as a {\em bingyin} day in two separate chapters.  
We have verified that this is in fact the case not only in the current standard Zhonghua edition of 
the {\em Old History of the Tang Dynasty}, but also in the very edition that Ho Peng Yoke cites.\footnote{See {\em Jiu 
Tang shu} in {\em Bona ben ershisishi (The Hundred Patches Edition of
the Standard Twenty-Four Histories)}, 820 vols., 1930-37, Vol. 365.11.20b and Vol. 372.36.10a.}
Hasegawa (1980) repeats Ho$^{\prime}$s (1962) error regarding the dating of the comet, 
despite having cited earlier catalogues that give the date as Jan 17 including Pingre (1784) 
and Williams (1871).

In addition to the aforementioned sources, Ho (1962) also cites an appearance of the 
comet in chapter 359 of the {\em Dai Nihon shi} ({\em Great History of Japan}) dated to the 
23rd day of the 12th month of the 3rd year of the Hoki reign period, 
AD 773 Jan 20, and as previously published by Kanda (1934, 1935).\footnote{
Hasegawa (1980) also cites Kanda (1935), but does not mention the Jan 20 date.}  
The Japanese observations are not inconsistent with the Chinese reports, but report a different date.  
This may be due to a true sighting on a different night, a few days later than the Chinese sighting.

\section{Result: A normal comet}

There are several parts of the translation in Liu et al (2014) that are unjustifiable in light of the historical texts on which it is supposed to be based.  
The opening phrase of the translation ({\em A comet collided with the Earth's atmosphere}) can only be described as an anachronistic interpolation.  
The te- xts never use any word meaning {\em atmosphere, collide}, nor {\em coma}.  
Moreover, while Liu et al. (2014) create the impression that a text from the Tang Dynasty described a coma that 
{\em stretched across the whole sky}, the earliest textual evidence to support such a claim in fact dates to AD 1060, 
nearly three hundred years after the event, and it speaks about a comet tail ({\em long star ... its length traversed the sky}), not a coma.  

As for the claim that the comet came {\em from the constellation Orion}, 
this is perhaps best dismissed as an infelicitous translation of the 
preposition {\em yu}, which sometimes does mean {\em from}, but here clearly means {\em in} or {\em at}.
Nor is there anything in any of the historical records to support the claim that the comet disappeared within one day.  
The texts give the date for the initial appearance of the comet, but do not specify the duration of time for which it was visible.
If the Japanese reports are credible, this indicates that the comet would have been visible for at least three days under clear conditions.

Pre-modern Chinese astronomy does not warrant the cl- aim that comets are located outside the Earth$^{\prime}$s
atmosphere or orbit the Sun (or the Earth). They were generally thought to be part of the Earth$^{\prime}$s atmosphere itself.
Hence, it would not have been imaginable for Chinese at that time, that a comet would {\em collide} with the Earth$^{\prime}$s atmosphere.

Finally, the phrase {\em yu tu}, which Liu et al. (2014) translate as {\em dust rain} and believe to refer to {\em cometary material} in 
the Earth$^{\prime}$s atmosphere, occurs no fewer than thirty-five times in the two Tang histories, 
where it is frequently associated with high winds and inclement weather, and means simply {\em dust storm.}
Of the six remaining references to {\em yu tu} in the {\em Old History of the Tang Dynasty}, for instance, 
two specify that the dust storms occurred in the context of heavy winds ({\em da feng}), 
while a third reference occurs in a chronicle of {\em thunderclaps and violent rainstorms} 
({\em leizhen baoyu}).\footnote{{\em Jiu Tang shu} 19B.710; 20A.779; 37.1362}
Three of the six references also specify the capital city ({\em jing shi}) as the location
where dust storms occurred.\footnote{{\em Jiu Tang shu} 37.1362; 7.145; 13.373.  
The final occurrence of {\em yu tu} gives a date when a dust storm occurred, 
but does not specify a location ({\em Jiu Tang shu} 3.43).}
None of these references mention any connection between {\em yu tu} and comets.
Dust storms and other meteorological phenomena, such as rainbows, oddly shaped clouds, and unseasonable weather, 
are included in Chinese treatises on astronomy, 
or more precisely, celestial patterns, because the early and medieval Chinese did not distinguish between meteorology and astronomy.

The claim that an event where a {\em comet collided with the Earth$^{\prime}$s atmosphere} would be {\em well established} (Liu et al. 2014) 
in the historical record is entirely unwarranted. 

We conclude that the Chinese just observed a more or less normal comet, possibly with an unusually long tail,
on (or beginning on) the night of AD 773 Jan 17 - after a day on which a dust storm occurred.
In Japan, the comet was also observed, probably on AD 773 Jan 20; therefore, it may have been visible for several nights.
If the comet was observed in Japan on AD 773 Jan 20, then it cannot have {\em collided with the Earth$^{\prime}$s atmosphere} before.

\acknowledgements
RN would like to thank Valeri Hambaryan for pointing us to the Liu et al. paper and
both Dagmar L. Neuh\"auser and Valeri Hambaryan for enlightening discussion about this event.
Finally, the authors would also like to thank the anonymous reviewer for several very helpful comments.

{}

\end{document}